\documentclass[12pt,thmsa]{article}
\usepackage{amssymb}

\input tcilatex
\QQQ{Language}{
American English
}

\begin{document}

\begin{center}
\textbf{MAGNETIC FIELD OF RELATIVISTIC NONLINEAR PLASMA WAVE}

\ 

Arsen G. Khachatryan

\textit{Yerevan Physics Institute, Alikhanian Brothers Street 2,}

\textit{Yerevan 375036, Armenia}

\textit{e}-\textit{mail: khachatr@moon.yerphi.am}
\end{center}

Longitudinal and transverse behavior of magnetic field of relativistic
nonlinear three-dimensional plasma wave is investigated. It is shown that
the magnetic field of the wave is different from zero and performs higher
frequency oscillations compared to the plasma electron frequency. An
increase in the nonlinearity leads to strengthening of magnetic field. The
oscillations of magnetic field in transverse direction arise, that caused by
the phase front curving of nonlinear plasma wave. The numerical results well
conform with predictions of the analytical consideration of weakly-nonlinear
case.

PACS number(s): 52.35.Mw, 52.40.Mj, 52.40.Nk

\ 

The progress in the technology of ultrahigh intensity lasers and high
current relativistic charged bunch sources permits the use of laser pulses$%
^1 $ or charged bunches$^2$ for excitation of strong plasma waves. The
excited plasma waves can be used, for example, for acceleration of charged
particles and focusing of bunches.$^2$ The amplitude of longitudinal
electric field $E_{\max }$ of relativistic plasma waves excited in cold
plasma is limited by the relativistic wave-breaking field$^3$ $%
E_{rel}=[2(\gamma -1)]^{1/2}E_{WB}/\beta $, where $\gamma =(1-\beta
^2)^{-1/2}$ is a relativistic factor, $\beta =v_{ph}/c$ is a dimensionless
phase velocity of the wave, $E_{WB}=m_e\omega _{pe}v_{ph}/e$ ($%
E_{WB}[V/cm]\approx 0.96n_p^{1/2}[cm^{-3}]$) is the conventional
nonrelativistic wave-breaking field, $\omega _{pe}=(4\pi n_pe^2/m_e)^{1/2}$
is the electron plasma frequency,\textbf{\ } $n_p$ is the equilibrium
density of plasma electrons, $m_e $ and $e$ are the mass and absolute value
of the electron charge.

The linear plasma wave theory is valid when $E_{\max }\ll E_{WB}$. It is
well known that a three-dimensional linear plasma wave in a cold plasma and
in the absence of external fields is potential.$^4$ In this case the
magnetic field of plasma wave is zero. The magnetic field is absent also in
one-dimensional nonlinear case due to the symmetry of the problem. However,
magnetic field of nonlinear three-dimensional plasma wave with the
relativistic phase velocity ($\beta \approx 1$) was not studied up to now
and our aim is to investigate this problem.

We shall study nonlinear plasma waves (wake waves) excited in cold plasma by
relativistic electron bunches or intense laser pulses (drivers) and suppose
the azimuthal symmetry of the problem. In this case for non-zero components
of plasma electrons momentum and electromagnetic field of the wave we have
the following set of equations$^{5,6}$:

\begin{equation}
\beta \frac{\partial P_z}{\partial z}-\frac{\partial \gamma _e}{\partial z}%
-\beta ^2E_z=0,  \tag{1}
\end{equation}

\begin{equation}
\beta \frac{\partial P_r}{\partial z}-\frac{\partial \gamma _e}{\partial r}%
-\beta ^2E_r=0,  \tag{2}
\end{equation}
\begin{equation}
-\frac{\partial H_\theta }{\partial z}+\beta \frac{\partial E_r}{\partial z}%
+\beta _rN_e=0,  \tag{3}
\end{equation}
\begin{equation}
\nabla _{\bot }H_\theta +\beta \frac{\partial E_z}{\partial z}+\beta
_zN_e+\beta \alpha =0,  \tag{4}
\end{equation}
\begin{equation}
\beta \frac{\partial H_\theta }{\partial z}-\frac{\partial E_r}{\partial z}+%
\frac{\partial E_z}{\partial r}=0,  \tag{5}
\end{equation}
\begin{equation}
N_e=1-\alpha -\nabla _{\bot }E_r-\frac{\partial E_z}{\partial z}.  \tag{6}
\end{equation}
As usual, the Eqs. (1) and (2) were derived taking into account that the
curl of the generalized momentum is zero, $\beta ^2\mathbf{H}-rot\mathbf{P}%
=0 $, or in our case 
\begin{equation}
\beta ^2H_\theta +\frac{\partial P_z}{\partial r}-\frac{\partial P_r}{%
\partial z}=0.  \tag{7}
\end{equation}
In Eqs. (1) - (6) $\gamma _e=(1+P_z^2+P_r^2+a^2/2)^{1/2}$ , $\beta _{z,\text{
}r}=P_{z,\text{ }r}$ $/\gamma _e$ and $N_e=n_e/n_p$ are respectively a
relativistic factor, dimensionless components of velocity and dimensionless
density of plasma electrons, $z=(\omega _{pe}/v_{ph})(Z-v_{ph}t),$ $\alpha
=n_b(z,r)/n_p$, $n_b$ is density of bunch electrons, $a=eE_0(z,r)/m_ec\omega
_0$, where $E_0$ and $\omega _0$ are the amplitude and frequency of laser
pulse, $\nabla _{\bot }=\partial /\partial r+1/r$. Also the following
dimensionless variables have been used: the space variables are normalized
on $\lambda _p/2\pi =v_{ph}/\omega _{pe}$, where $\lambda _p$ is the linear
plasma wavelength, the momenta and velocities - respectively on $m_ec$ and
the velocity of light and the strengths of electric and magnetic fields - on
the nonrelativistic wave-breaking field $E_{WB}$. In the general case the
analytical consideration of the problem seems impossible. First we consider
weakly-nonlinear case and then present numerical results.

In the weakly-nonlinear case we suppose $\beta =1$ and use the
generalization of the well known expansion$^7$ which was used to study
one-dimensional nonlinear relativistic-plasma-waves$^8$:

\begin{eqnarray}
u(r,z) &=&\varepsilon u_1(r,\Psi )+\varepsilon ^2u_2(r,\Psi )+\varepsilon
^3u_3(r,\Psi )+...,  \tag{8} \\
\partial \Psi /\partial z &=&1+\varepsilon k_1(r)+\varepsilon ^2k_2(r)+..., 
\nonumber
\end{eqnarray}
where $u$ stands for normalized values $P_{z,r}$, $E_{z,r}$ or $H_\theta $, $%
\varepsilon =E_z^{\max }\ll 1$ is the small parameter, $P_{z,r}$, $E_{z,r}$, 
$H_\theta \ll 1$, $\Lambda _p$ is the nonlinear plasma wavelength.
Substituting expansion (8) into equations (1)-(5) we have:

\begin{equation}
\frac{\partial P_{zi}}{\partial \Psi }-E_{zi}=S_{1,i},  \tag{9}
\end{equation}

\begin{equation}
\frac{\partial P_{ri}}{\partial \Psi }-E_{ri}=S_{2,i}  \tag{10}
\end{equation}

\begin{equation}
-\frac{\partial H_{\theta i}}{\partial \Psi }+\frac{\partial E_{ri}}{%
\partial \Psi }+P_{ri}=S_{3,i},  \tag{11}
\end{equation}

\begin{equation}
\nabla _{\bot }H_{\theta i}+\frac{\partial E_{zi}}{\partial \Psi }%
+P_{zi}=S_{4,i},  \tag{12}
\end{equation}

\begin{equation}
\frac{\partial H_{\theta i}}{\partial \Psi }-\frac{\partial E_{ri}}{\partial
\Psi }+\frac{\partial E_{zi}}{\partial \Psi }=S_{5,i},  \tag{13}
\end{equation}
where subscript $i$ denotes the order of approximation and $S_{1-5,i}$ are
the nonlinear functions of $u_{i-1}$, $u_{i-2}$, $...$, and $u_1$. In the
linear case ($i=1$) $S_{1-5,1}=0$ and one can obtain well known solutions
(see, e.g., Ref. 9): $P_{z1}=-R\sin \Psi ,$ $P_{r1}=(dR/dr)\cos \Psi ,$ $%
E_{z1}=-R\cos \Psi ,$ $E_{r1}=-(dR/dr)\sin \Psi ,$ and $H_{\theta 1}=0$,
where $R(r)$ depends on the radial distribution in the exciting source, i.
e., on $\alpha (r)$ or $a^2(r)$. In the second approximation, from the
condition of absence of resonance terms (proportional to $\sin \Psi $ or $%
\cos \Psi $) one can find that $k_1=0$. Then

\begin{eqnarray*}
S_{1,2} &=&(1/2)\partial (P_{r1}^2+P_{z1}^2)/\partial \Psi , \\
S_{2,2} &=&(1/2)\partial (P_{r1}^2+P_{z1}^2)/\partial r, \\
S_{3,2} &=&P_{r1}(\nabla _{\bot }E_{r1}+\partial E_{z1}/\partial \Psi ), \\
S_{4,2} &=&P_{z1}(\nabla _{\bot }E_{r1}+\partial E_{z1}/\partial \Psi ), \\
S_{5,2} &=&0,
\end{eqnarray*}
and for magnetic field strength we obtain the following equation:

\begin{equation}
\lbrack (\partial /\partial r)\nabla _{\bot }-1]H_{\theta 2}=\partial
S_{4,2}/\partial r-\partial S_{3,2}/\partial \Psi .  \tag{14}
\end{equation}
Solution of this equation is

\begin{equation}
H_{\theta 2}=f_1(r)+f_2(r)\cos (2\Psi ),  \tag{15}
\end{equation}
where $f_{1,2}(r)$ satisfies to equations

\begin{eqnarray*}
(\Delta _{\bot }-r^{-2}-1)f_1 &=&(1/2)(d/dr)[R(\Delta _{\bot }-1)R], \\
(\Delta _{\bot }-r^{-2}-1)f_2 &=&(1/2)[(dR/dr)\Delta _{\bot }R-R(d/dr)\Delta
_{\bot }R],
\end{eqnarray*}
here $\Delta _{\bot }=\nabla _{\bot }(d/dr)$ is the transverse part of
Laplacian. In the $i-th$ approximation $H_{\theta i}\sim \cos (i\Psi )$. The
dependence $k_2(r)$ can be obtained in the third approximation. So, the
weakly-nonlinear theory predicts that the magnetic field of 3D nonlinear
plasma wave is not zero, in opposite to the 3D linear case and to 1D
nonlinear one, and oscillates at the harmonics of the linear plasma
frequency; the nonlinear wavelength changes in the radial direction.

We have solved Eqs. (1) - (6) numerically choosing the Gaussian profile of
the driver both in longitudinal and transverse directions:

\begin{equation}
A(z,r)=A_0\exp [-(z-z_0)^2/\sigma _z^2]\exp (-r^2/\sigma _r^2),  \tag{16}
\end{equation}
where $A(z,r)$ stands for $\alpha $ or $a^2$. Fig. 1 shows three-dimensional
linear plasma wave excited by relativistic electron bunch. The linear
numerical solution obtained well agreed with predictions of the linear
theory.$^2$ As is seen in Figure 1(b), in the linear case the magnetic field
excited in plasma is localized in the range occupied by the bunch and in the
wake $E_z^{\max }\gg H_\theta ^{\max }\thickapprox 0$, that correspond to
the linear theory.

In the nonlinear regime the behavior of plasma waves is qualitatively
changed. In Fig. 2 one can see a nonlinear plasma wave that is excited by an
intense laser pulse. The main difference here from the linear case is the
change of shape and length of the wave with the radial coordinate $r$ [see
Fig. 2(a)]. The magnetic field strength in the nonlinear plasma wave as
shown in Figure 2(b) is different from zero and has the magnitude comparable
with that of other components of the field. The magnitude of higher
frequency oscillations (as compared to the plasma frequency) performed by
the magnetic field along $z$ grows in proportion to the nonlinearity. Such a
behavior of magnetic field is a purely nonlinear effect. Indeed, in the
linear case $H_\theta =0$, i.e., the contribution of momentum components at
the plasma frequency in the expression (7) is compensated. The nonlinearity
of the wave implies a rise of higher harmonics in $P_z$ and $P_r$. According
to (7), the rise of magnetic field is due to these harmonics and this
accounts for frequent oscillations seen in Figure 2(b), that conforms with
the weakly-nonlinear analytical consideration presented above. On the other
hand, the non-zero magnetic field means, according to (7), that the motion
of plasma electrons in the nonlinear wave is turbulent ($rot\mathbf{P}\neq 0$%
). The degree of turbulence (the measure of which is $H_\theta $) grows in
proportion to the nonlinearity.

It is easy to see that due to the dependence of the wavelength on$\,\,\,r$,
the field in the radial direction grows more chaotic as the distance from
the driver increases. In fact, the oscillations of plasma for different $%
\,r\,\,$are ''started'' behind the driver with nearly equal phases but
different wavelengths [see Fig. 2(a)]. As $\left| z\right| $\thinspace
\thinspace \thinspace increases, the change of phase in transverse direction
(for fixed $z$) becomes more and more marked. This leads to a curving of the
phase front and to ''oscillations'' in the transverse direction.$^{6,10}$
The radial behavior of the longitudinal electric field and magnetic field of
the nonlinear plasma wave is presented in Fig. 3. Qualitatively the radial
dependence of the field differs from that of the linear case by the change
of sign and ''steepening'' of fields along $r$ and that is connected with
the curvature of wave phase front.

So, we have found that the magnetic field of nonlinear plasma wave is
different from zero and performs higher frequency oscillations as compared
to the plasma frequency. The latter qualitatively differs from the linear
case and one-dimensional nonlinear one. The obtained numerical results well
agreed with the predictions of weakly-nonlinear theory.

\begin{center}
\textbf{REFERENCES}
\end{center}

$^1$T. Tajima and J. M. Dawson, Phys. Rev. Lett. \textbf{43}, 267 (1979).

$^2$R. D. Ruth, A. W. Chao, P. L. Morton, and P. B. Wilson, Part. Accel. 
\textbf{17}, 171 (1985); P. Chen, Part. Accel. \textbf{20}, 171 (1987).

$^3$A. I. Akhiezer and R .V. Polovin, Zh. Eksp. Teor. Fiz. \textbf{30}, 915
(1956) [Sov. Phys. JETP \textbf{3}, 696 (1956)].

$^4$N. A. Krall and A. W. Trivelpice, \textit{Principles of Plasma Physics}
(McGraw-Hill Book Co., NY, 1973).

$^5$B. N. Breizman, T. Tajima, D. L. Fisher, and P. Z. Chebotaev, In: 
\textit{Research Trends in Physics: Coherent Radiation and Particle
Acceleration , edited by A. Prokhorov }(American Institute of Physics, New
York, 1992), pp. 263-287; K. V. Lotov, Phys. Plasmas \textbf{5}, 785 (1998).

$^6$A. G. Khachatryan and S. S. Elbakian. Two-dimensional nonlinear regime
in the Plasma Wakefield Accelerator, Proceedings PAC'99, edited by A.Luccio
and W.MacKay (IEEE, Piscataway, NJ, 1999), pp. 3663-3665.

$^7$G. B. Witham, \textit{Linear and Nonlinear Waves} (Wiley, New York,
1974), Chap. 13.

$^8$A. G. Khachatryan, Phys. Rev. E\textit{\ }\textbf{58}, 7799 (1998).

$^9$R. Keinigs and M. E. Jones, Phys. Fluids \textbf{30}, 252 (1987); A. G.
Khachatryan, A. Ts. Amatuni, S. S. Elbakian, and E. V. Sekhpossian, Plasma
Phys. Rep. \textbf{22}, 576 (1996).

$^{10}$A. G. Khachatryan, Phys. Rev. E\textit{\ }\textbf{60}, 6210 (1999).

\begin{center}
\textbf{FIGURE CAPTIONS}
\end{center}

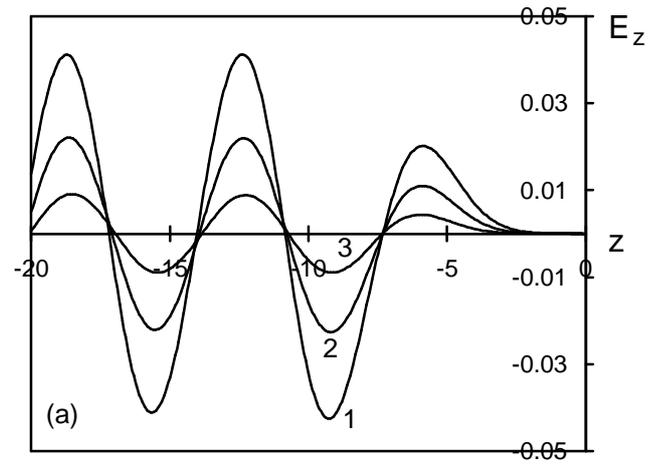
Figure 1. The linear plasma wave excited by an electron bunch with gaussian
profile;$\,\,\alpha _0=0.1$, $\sigma _z=2$,$\,\,\sigma _r=0.5$, $\gamma =10$%
. (a). The dimensionless strength of longitudinal electric field. 1---$%
\,E_z(z)$ at the axis, $r=0$; 2 --- $r=1$; 3 --- $r=2$. (b). The strength of
azimuthal magnetic field. 1 --- $r=0.2$; 2 --- $r=0.75$; 3 --- $r=2$.

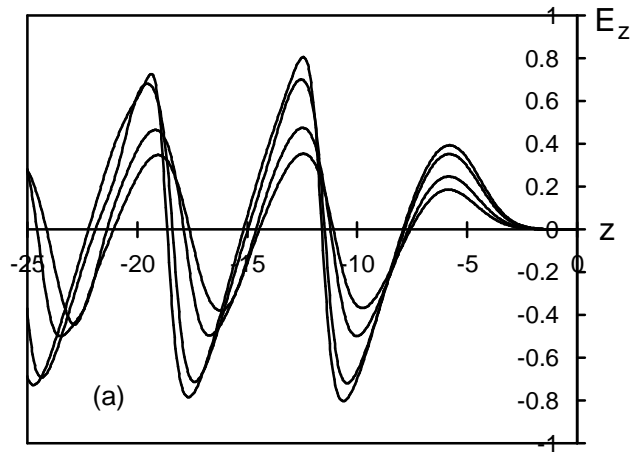
Figure 2. The nonlinear plasma wave excited by laser pulse. The pulse
parameters are : $a_0^2=3.6$, $\sigma _z=2$,$\,\,\sigma _r=5$, $\gamma =10$.
(a). The longitudinal electric field $E_z$. $r=0,2,4$ and $5$ in the order
of magnitude reduction. (b). The magnetic field strength. 1 --- $r=2$; 2 --- 
$r=4$.

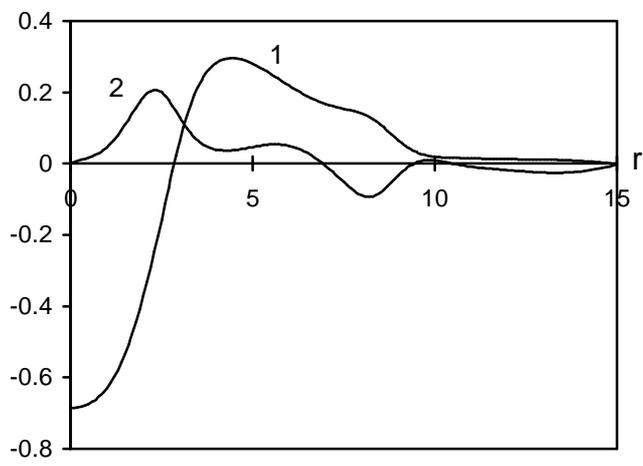
Figure 3. The radial behavior of field for the case given in Figure 2 for $%
z=-25.$ 1 --- longitudinal electric field $E_z(z=-25,r)$; 2 --- magnetic
field $H_\theta (z=-25,r)$.

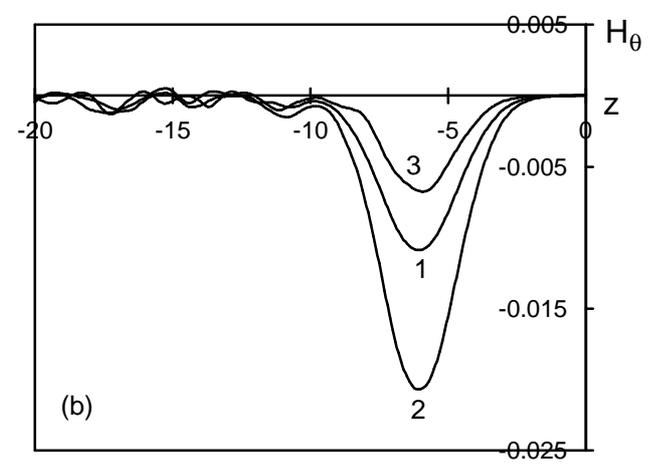
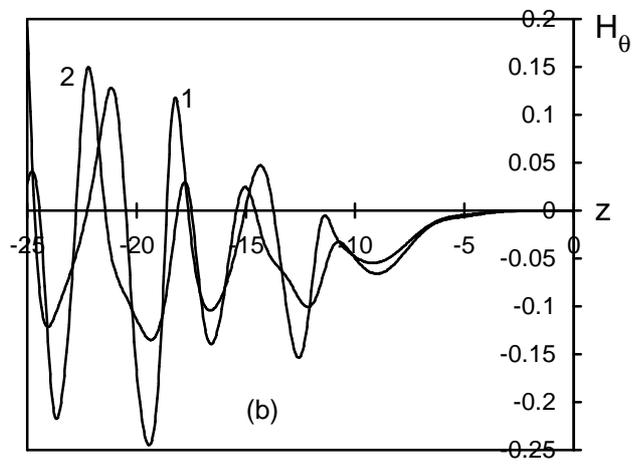
\end{document}